\documentclass[aps,prd,superscriptaddress,amssymb,nofootinbib]{revtex4}
\usepackage{graphicx} 
\usepackage{color}
\def\rts{\sqrt s}
\def\mx{M_X}
\def\ie{{\it i.e.}}

\def\lsim{\mathrel{\raise.3ex\hbox{$<$\kern-.75em\lower1ex\hbox{$\sim$}}}}
\def\gsim{\mathrel{\raise.3ex\hbox{$>$\kern-.75em\lower1ex\hbox{$\sim$}}}}
\def\mh{m_h}
\def\rn{r_{\vec n}}
\def\hprime{h'}
\def\mhprime{m_{h'}}
\def\ie{{\it i.e.}}

\def\gkl{g_{\vec k\vec l}}

\def\what{\widehat}
\def\kap{\kappa}
\def\calo{{\cal O}}
\def\cala{{\cal A}}

\def\beq{\begin{equation}}
\def\eeq{\end{equation}}
\def\rhodel{\rho_\del}

\def\gaminv{\Gamma_{\rm inv}}
\def\gamh{\Gamma_h^{SM}}
\def\gamres{\Gamma_{\rm res}}
\def\rhoinv{\rho_{\rm inv}}
\def\tev{~{\rm TeV}}
\def\gev{~{\rm GeV}}

\def\md{M_D}
\def\epem{e^+e^-}

\def\mhprime{m_{h'}}
\def\mw{m_W}
\def\to{\rightarrow}
\def\ptl{\partial}
\def\beq{\begin{equation}}
\def\eeq{\end{equation}}
\def\bea{\begin{eqnarray}}
\def\eea{\end{eqnarray}}
\def\nn{\nonumber}
\def\half{{1\over 2}}

\def\rhalf{{1\over \sqrt 2}}
\def\calo{{\cal O}}
\def\call{{\cal L}}

\def\del{\delta}
\def\eps{\epsilon}
\def\lam{\lambda}
\def\anti{\overline}
\def\delfac{\sqrt{{2(\del-1)\over 3(\del+2)}}}
\def\heff{h'}
\def\square{\boxxit{0.4pt}{\fillboxx{7pt}{7pt}}\hspace*{1pt}}
    \def\boxxit#1#2{\vbox{\hrule height #1 \hbox {\vrule width #1
             \vbox{#2}\vrule width #1 }\hrule height #1 } }
    \def\fillboxx#1#2{\hbox to #1{\vbox to #2{\vfil}\hfil}   }

\def\gev{~{\rm GeV}}
\def\gam{\gamma}
\def\sn{s_{\vec n}}

\def\sk{s_{\vec k}}
\def\sls{s_{\vec l}}
\def\sm{s_{\vec m}}

\def\mn{m_{\vec n}}

\def\mh{m_h}
\def\sumn{\sum_{\vec n>0}}
\def\summ{\sum_{\vec m>0}}
\def\suml{\sum_{\vec l>0}}
\def\sumk{\sum_{\vec k>0}}
\def\vl{\vec l}
\def\vk{\vec k}
\def\ml{m_{\vl}}
\def\mk{m_{\vk}}
\def\mpl{M_P}
\def\to{\rightarrow}
\def\ptl{\partial}
\def\beq{\begin{equation}}
\def\eeq{\end{equation}}
\def\bea{\begin{eqnarray}}
\def\eea{\end{eqnarray}}
\def\nn{\nonumber}
\def\half{{1\over 2}}
\def\rhalf{{1\over \sqrt 2}}
\def\calo{{\cal O}}
\def\call{{\cal L}}

\def\del{\delta}
\def\eps{\epsilon}
\def\lam{\lambda}
\def\anti{\overline}
\def\delfac{\sqrt{{2(\del-1)\over 3(\del+2)}}}
\def\heff{h'}
\def\square{\boxxit{0.4pt}{\fillboxx{7pt}{7pt}}\hspace*{1pt}}
    \def\boxxit#1#2{\vbox{\hrule height #1 \hbox {\vrule width #1
             \vbox{#2}\vrule width #1 }\hrule height #1 } }
    \def\fillboxx#1#2{\hbox to #1{\vbox to #2{\vfil}\hfil}   }

\def\gev{~{\rm GeV}}

\def\gam{\gamma}
\def\sn{s_{\vec n}}
\def\sm{s_{\vec m}}

\def\mn{m_{\vec n}}
\def\mh{m_h}
\def\sumn{\sum_{\vec n>0}}
\def\summ{\sum_{\vec m>0}}
\def\vl{\vec l}
\def\vk{\vec k}
\def\ml{m_{\vl}}
\def\mk{m_{\vk}}
\def\gaminv{\Gamma_{\rm inv}}

\begin{document}
\title{INVISIBLE HIGGS DECAYS FROM HIGGS GRAVISCALAR MIXING}

\author{Daniele Dominici}
\email[]{dominici@fi.infn.it}

\affiliation{Department of Physics, University of Florence, Sesto
  F. (FI), Florence and Istituto Nazionale di Fisica Nucleare,
Sezione di Firenze, Italy}

\author{John F. Gunion}
\email[]{gunion@physics.ucdavis.edu}

\affiliation{Department of Physics, University of California at Davis, Davis, CA 95616}

\date{\today}
\begin{abstract}
  We recompute the invisible Higgs decay width arising from
  Higgs-graviscalar mixing in the ADD model, comparing the original
  derivation in the non-diagonal mass basis to that in a diagonal mass
  basis. The results obtained are identical (and differ by a factor of
  2 from the original calculation) but the diagonal-basis
  derivation is pedagogically useful for clarifying the physics of the
  invisible width from mixing. We emphasize that both derivations make
  it clear that a direct scan in energy for a process such as $WW\to
  WW$ mediated by Higgs plus graviscalar intermediate resonances would
  follow a {\it single} Breit-Wigner form with total width given by
  $\Gamma^{tot}=\Gamma_h^{SM}+\Gamma_{invisible}$. We also compute the
  additional contributions to the invisible width due to direct Higgs
  to graviscalar pair decays. We find that the invisible width due to
  the latter is relatively small unless the Higgs mass is comparable
  to or larger than the effective extra-dimensional Planck mass.

\end{abstract}

\maketitle

\section{INTRODUCTION}
In several extensions of the Standard Model (SM) there exist
mechanisms which modify the Higgs decay rates in channels observable
at the LHC.  One recent example is the Randall Sundrum
model~\cite{Randall:1999ee} where the Higgs-radion mixing can modify
Higgs production and decay at the
LHC~\cite{Dominici:2002jv,Hewett:2002nk}.  These effects may be
detected both through a reduction in the Higgs yield and in the direct
observation of radion decays
\cite{Dominici:2002jv,Hewett:2002nk,Battaglia:2003gb}.  There are also
examples where the reduction comes from a substantial invisible width,
as occurs for example in those supersymmetric models in which the
Higgs has a large branching ratio into the lightest gravitinos or
neutralinos.  Invisible decay of the Higgs is also predicted in models
with large extra dimensions felt by gravity (ADD)
\cite{Arkani-Hamed:1998rs,Antoniadis:1998ig}, our focus in this paper.

In ADD models, the
presence of an interaction between the Higgs $H$ and the Ricci scalar
curvature of the induced 4-dimensional metric, $g_{ind}$,
given by the action \cite{Giudice:2000av}
\beq
S=-\xi \int d^4 x \sqrt{g_{ind}}R(g_{ind})H^\dagger H
\eeq
generates, after the usual shift $H=((v+h)/\sqrt{2},0)$, the mixing term 
\begin{equation}
{\cal L}_{mix}= - \frac {2\xi v M_H^2}{\mpl} 
\sqrt {\frac {3(\delta -1)}
{\delta +2}}
h \Sigma_{\vec n} \phi_G^{\vec n}
\end{equation}
where the $\phi_G^{\vec n}$ are complex graviscalar fields, $\mpl$ is the reduced
Planck mass  ($\mpl=(8\pi G_N)^{-1/2}$),
 $\delta$ is the number of extra dimensions and $\xi$ is a dimensionless
 parameter.
Noting that hermiticity requires  $\phi_G^{\vec n}=[\phi_G^{-\vec n}]^*$
and 
writing $\phi_G^{\vec n}={1\over\sqrt 2}(s_{\vec n}+i a_{\vec n})$, we
may 
restrict  the sums to $\vec n>0$, by which we mean the first non-zero entry
of $\vec n$ is positive.\footnote{
It is quite crucial to explicitly keep only $\vec n>0$ states since $s_{\vec n<0}$ is
not independent of $s_{\vec n>0}$.  This is especially important
for obtaining correct Feynman rules that avoid double counting.
} Then, defining
\beq
\eps\equiv -{2\sqrt 2\over \mpl}\xi v \mh^2\sqrt{{3(\del-1)\over \del+2}}\,,
\label{epsdef}
\eeq
one obtains
\begin{equation}
{\cal L}_{\rm mix}=\epsilon  h \sum_{\vec n >0}s_{\vec n}\,,
\label{lmixform}
\end{equation}
where 
$s_{\vec n}$ is a CP-even canonically normalized graviscalar KK excitation with
mass $m_{\vec n}^2=4\pi^2 \vec n^2/L^2$, $L$ being the
size of each of the extra dimensions.

As a result of the above mixing, instead of a single Higgs boson, one
must consider the production of the full set of densely spaced mass
eigenstates all of which are mixing with one another.  The new
signature that arises as a result of this mixing is that the Higgs
boson will effectively acquire a possibly very large branching ratio
to invisible final states composed primarily of graviscalars. The
purpose of this paper is to first rederive the result obtained in
\cite{Giudice:2000av,Wells:2002gq} comparing the non-diagonal mass and
diagonal mass bases in the direct Feynman diagram approach. Our
approach clarifies the nature of this effect and also reveals a factor
two error in the original derivation (as confirmed in
\cite{Datta:2004jg}).  (A brief summary of our results and related
phenomenology appeared in \cite{Battaglia:2004js}.) These computations
set the stage for our main goal of computing the Higgs to graviscalar
pair width, in particular making it clear that such 
computations are most easily performed using the non-diagonal
(Lagrangian) basis states rather than the mass eigenstate basis. The
graviscalar pair states add to the invisible width coming purely from
Higgs-graviscalar mixing. As we show, this additional invisible width
is small relative to the mixing width if $\mh$ is small compared to
$\md$ ($\md$ is related
to the $D$ dimensional reduced Planck constant ${\overline M}_D$ by
$\md=(2\pi)^{\delta/(2+\delta)}{\overline M}_D$) but should be
accounted for in any eventual precision comparison between theory and
experiment if $\mh$ is comparable to or larger than $\md$.

\section{INVISIBLE WIDTH}

 In \cite{Giudice:2000av,Wells:2002gq} the invisible Higgs width is
calculated by extracting the imaginary part of the Higgs self energy,
including the effects of Higgs-graviscalar mixing. The result we
obtain following this general approach differs by a factor of two from
that of   \cite{Giudice:2000av,Wells:2002gq} (basically because of the
need to use properly normalized $\sn$ states)
and is given by 
\begin{eqnarray}
\Gamma_{inv}(h\to s_{\vec n})&=&
2\pi\xi^2 v^2 \frac {3(\delta -1)}
{\delta +2}
\frac {\mh^{1+\delta}}{\md^{2+\delta}}{S_{\delta -1}}
\sim (16\,MeV) 20^{2-\delta } \xi^2S_{\delta-1}\frac {3(\delta -1)}
{\delta +2} \left ( \frac {\mh}{150\, GeV} \right )^{1+\delta}
\left ( \frac 
{3\, TeV} {\md}\right )^{2+\delta}
\end{eqnarray}
where $S_{\delta-1}=2\pi^{\delta/2}/\Gamma(\delta/2)$
 denotes the surface of a unit radius sphere in $\delta$ dimensions.
 In this paper, we first repeat the derivation
of this result in the $h$ -- $s_{\vec n}$ basis,  \ie\ before
mass diagonalization. In our second derivation, we first diagonalize the Hamiltonian to
obtain the mass eigenstates. In both cases, we compute, by way of
example, the amplitude for $WW\to WW$ scattering coming from summing
over the diagonal Higgs and graviscalar eigenstate exchanges. The
derivations make it absolutely clear that a scan of the cross section
for $WW\to WW$ scattering over $s_{WW}$
would reveal a simple Breit-Wigner of width
$\Gamma_h^{SM}+\Gamma_{invisible}$, implying that a direct scan in
$s_{WW}$ can be used to determine $\gaminv$.  Further, after
integrating over energy, the invisible width suppresses the LHC Higgs rate
in the standard observable channels (such as $WW$) by a factor of $1/(1+R)$ where
\beq
R\equiv {\Gamma_{invisible}\over \Gamma_h^{SM}}
\label{rdef}
\eeq
can be quite substantial even for a Higgs boson with mass above the
$WW$ decay threshold. These two different ways of determining $R$ can
then be checked for consistency. In contrast, as pointed out in
\cite{Giudice:2000av}, a process such as $\epem\to Z^*\to Z +X$ cannot
be directly employed to determine $R$ by simply measuring the ratio of
the $X=invisible$ rate relative to the $X=visible$ rate.

\subsection{Derivation of the invisible width from graviscalar
  insertions into the Higgs propagator}

It is useful to first present a derivation of the above expression for
the invisible Higgs width following a procedure that is essentially
that of Refs.~\cite{Giudice:2000av,Wells:2002gq}. We begin with the
expression for the mixing Lagrangian given in Eq.~(\ref{lmixform}). We
consider a process such as $WW\to WW$ and recall that only the $h$
states have significant (\ie\ not suppressed by $1/\mpl$) coupling
to $WW$.  The contributing Feynman diagrams are such that one begins
with a $WW\to h$ vertex and ends with an $h\to WW$ vertex. There are
then diagrams with no $\sumn \sn$ insertions, one $\sumn \sn$ insertion
with two mixing vertices and so forth, resulting in a geometric series
that can be resummed to give an effective $WW\to WW$ $s$-channel
scattering amplitude with exactly the same form {\it including normalization} as a single Higgs
exchange but with an additional contribution to the self-energy of the
Higgs such that 
\beq 
\cala_{WW\to WW}={g_{WWh}^2 \over s_{WW}-\mh^2+i\mh\gamh}\to
{g_{WWh}^2 \over s_{WW}-\mh^2+i\mh\gamh+\Sigma(s_{WW})} 
\label{awwform}
\eeq 
where
\bea 
Im \Sigma(s)&=& -\epsilon^2 Im \sum_{\vec n>0}\frac 1
{s-m_n^2+i\eps} \rightarrow -\epsilon^2 \frac 1 4 \frac
{M_P^2}{\md^{2+\delta}}S_{\delta -1}(-\pi)s^{(\delta-2)/2} =2\pi
{3(\del-1)\over \del+2}\xi^2 v^2 m_h^2 \frac
{s^{(\delta-2)/2}}{\md^{2+\delta}}S_{\delta -1}\,.
\label{imsig}
\eea
This is interpreted as saying that the Higgs has acquired an
additional width given by 
\beq
\Gamma_{invisible}=\frac 1 {m_h}Im \Sigma(m_h^2)
=2\pi {3(\del-1)\over \del+2}\xi^2 v^2
\frac{m_h^{1+\delta}}{\md^{2+\delta}}S_{\delta -1}
\label{gaminvdef}
\eeq
that is deemed an invisible width since the $\sn$ graviscalar states
do not interact with ordinary matter and would be invisible in a
detector. 
 
The procedure for deriving the expression following the arrow in
Eq.~(\ref{imsig}) is as follows.  First, one converts the sum over
$\vec n>0$ to an integral over a
continuous spectrum of $s_{\vec n}$ masses as follows:
\bea
\sum_{\vec n>0}\frac 1 {s-m_n^2+i\eps}\rightarrow&&
\frac 1 2 \int dm^2\rho_\delta(m)\frac 1 {s-m^2+i\eps}
=\frac 1 4 \frac {M_P^2}{\md^{2+\delta}}S_{\delta -1}\int dm^2
\frac {m^{\delta-2}} {s-m^2+i\eps}
\eea
where the factor of $1/2$ arises in going from $\sum_{\vec n>0}$ to
$\sum_{\vec n}$ and 
we have used the following expression for the state density:
\beq
\rho_\delta(m)=\frac {L^\delta m^{\delta-2}}{(4\pi)^{\delta/2}\Gamma(\delta/2)}
=\frac {M_P^2}{\md^{2+\delta}}
\frac{\pi^{\delta/2}}{\Gamma(\delta/2)}
m^{\delta-2}=\half \frac {M_P^2}{\md^{2+\delta}}S_{\del-1}m^{\delta-2}
\label{rhodeldef}
\eeq
with $L^\delta=(2\pi)^\delta  \frac {M_P^2}{\md^{2+\delta}}$. 
Then, using 
\beq
Im \frac 1 {s-m^2+i\eps}=-\pi \delta(s-m^2)
\eeq
one has
\beq
Im \sum_{\vec n>0}\frac 1 {s-m_n^2+i\eps}\rightarrow
\frac 1 4 \frac {M_P^2}{\md^{2+\delta}}S_{\delta -1}(-\pi)s^{(\delta-2)/2}\,.
\eeq

We emphasize that Eq.~(\ref{awwform}) implies that a scan of the cross
section for $WW\to WW$ as  $s_{WW}$ is varied would reveal a single
Breit-Wigner of width $\gamh+\Gamma_{invisible}$ to the extent that
the real part of $\Sigma(s_{WW})$ can be neglected. However, 
the real part of $\Sigma(s)$ can lead to mass and wave
function renormalization (see Appendix), which corrections are of order 
$\mh^4/\md^4$. As a result, the apparent magnitude of the total width
measured in a scan will receive corrections of this order. In
addition, the interpretation of the normalization of
$\cala_{WW\to WW}$ (or any other process beginning with SM particles
and ending with SM particles) and of the effective pole location will
be similarly affected. In addition, our focus here, we find that there are further
corrections to the effective width coming from additional contributions
to $Im \Sigma$.  In particular, we will discuss the $h\to \sn\sm$ type
of insertions.  These we will find to be of order
$\xi^2(\mh/\md)^{2+\del}$ relative to the mixing width, 
and therefore potentially significant if $\mh>\md$.  However, in
practice the invisible width from the $\sn\sm$ pair final states is
suppressed sufficiently by two-body phase space that the pair-width to
mixing-width ratio is typically very small.

\begin{figure}[htbp]
\begin{center}
\includegraphics[width=8.0cm,height=8.0cm]{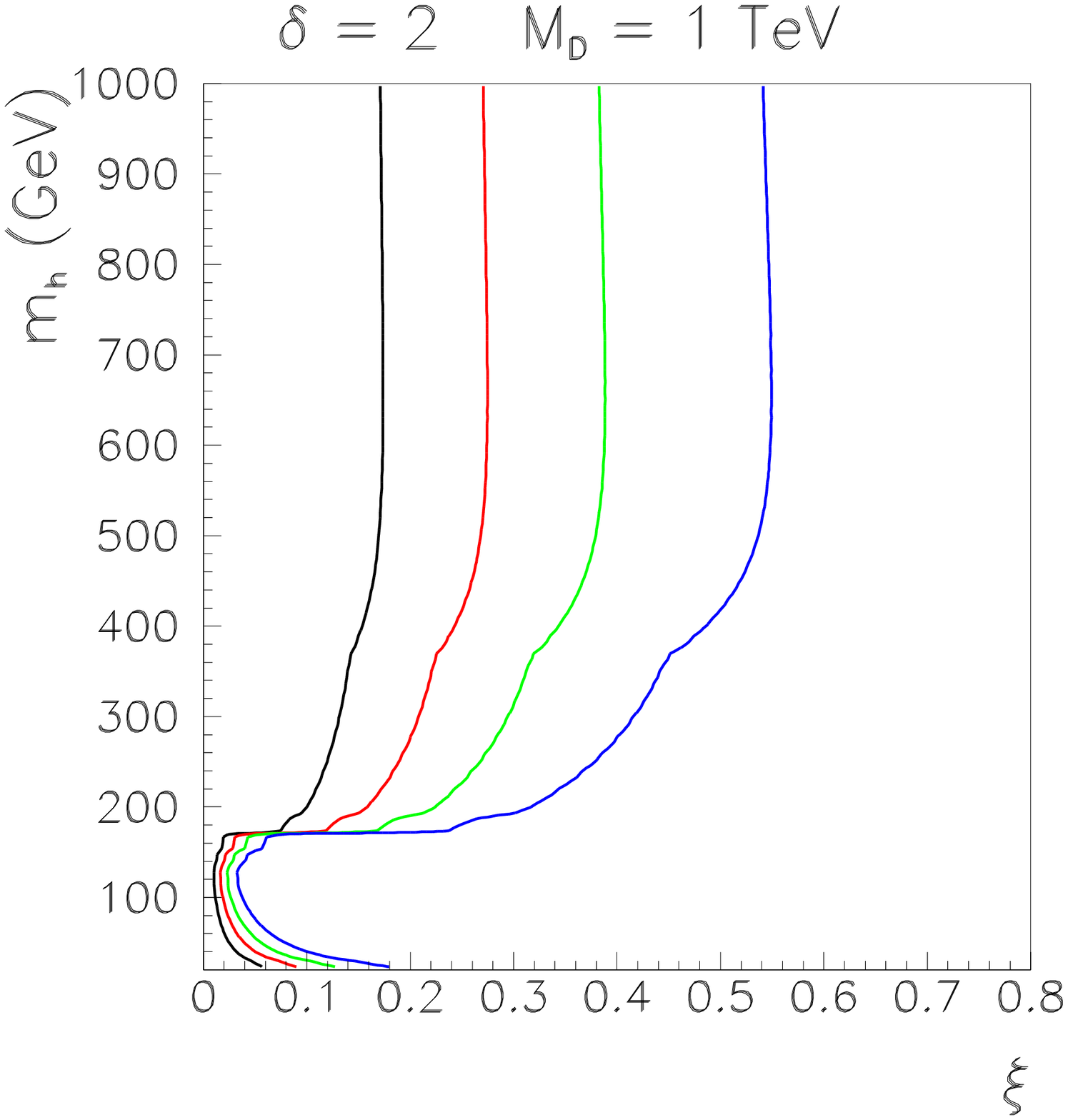} 
\includegraphics[width=8.0cm,height=8.0cm]{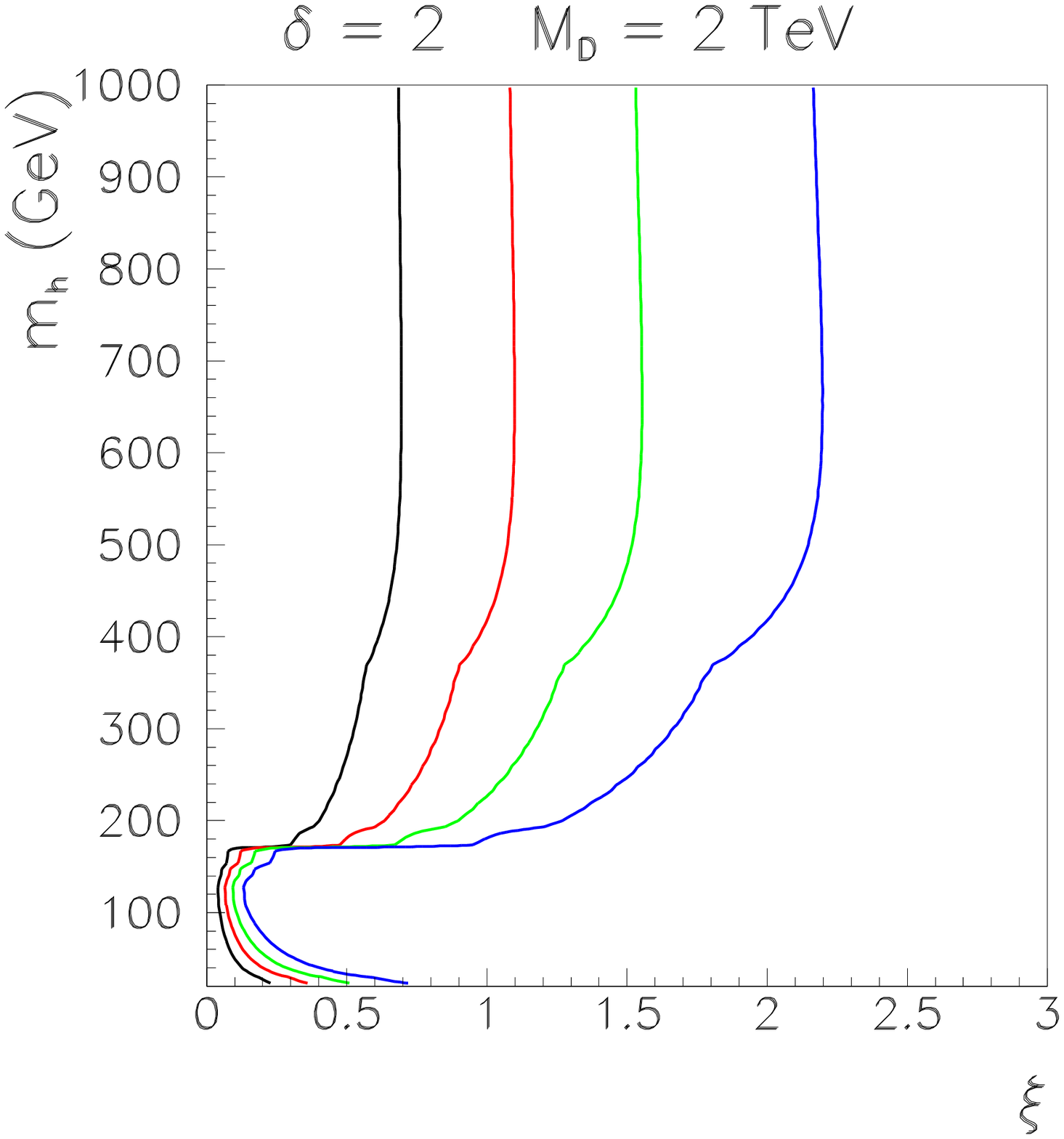}
\caption{We display contours of $R=0.1$, 0.25, 0.5 and 1 (from left to
  right) in the
  $m_h$--$\xi$ plane, for (left plot) $\del=2$ and $\md=1\tev$ and
  (right plot)
  $\del=2$ and $\md=2\tev$.}
\label{Rplot1}
\end{center}
\end{figure}

\begin{figure}[htbp]
\begin{center}
\includegraphics[width=8.0cm,height=8.0cm]{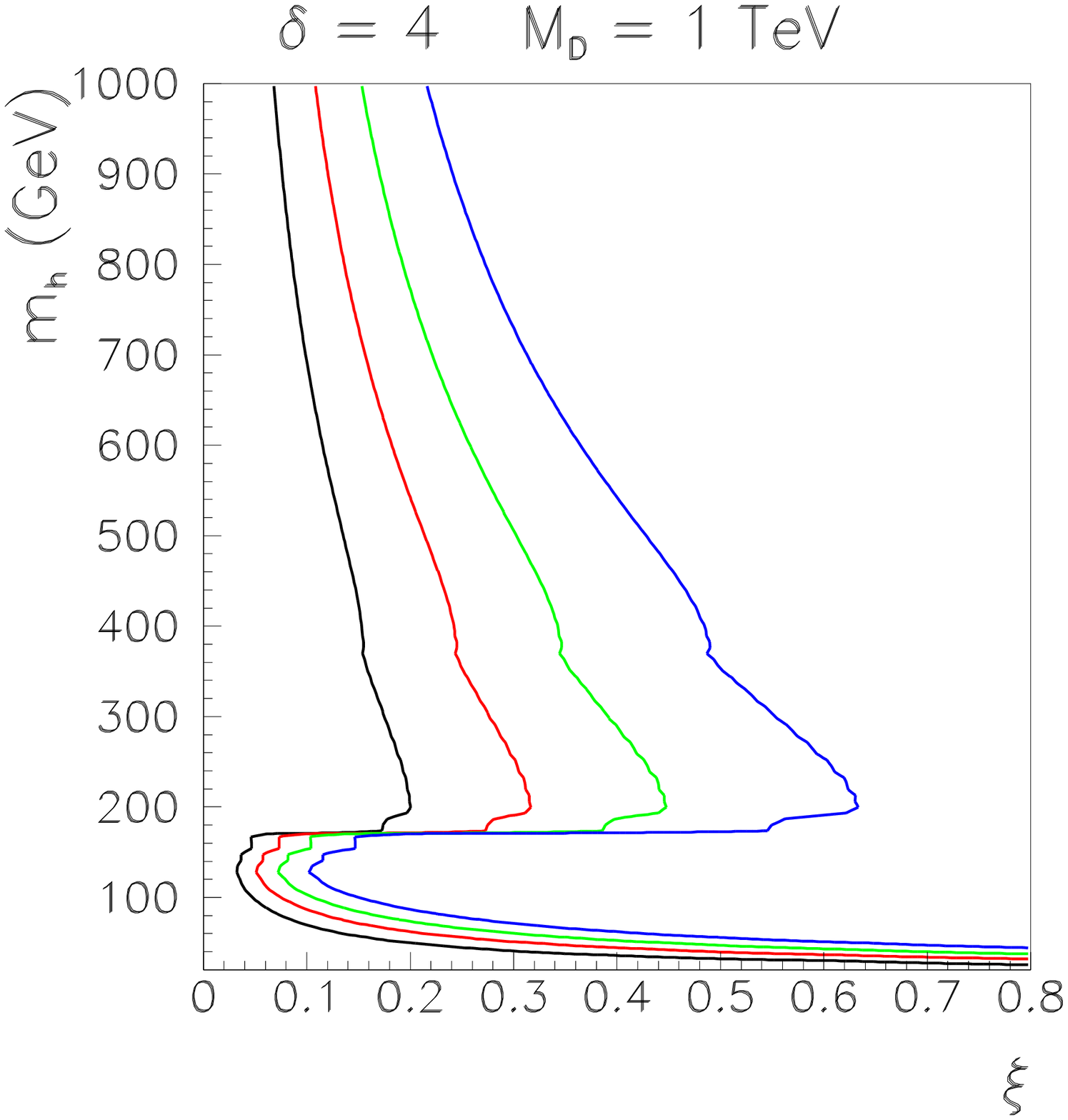} 
\includegraphics[width=8.0cm,height=8.0cm]{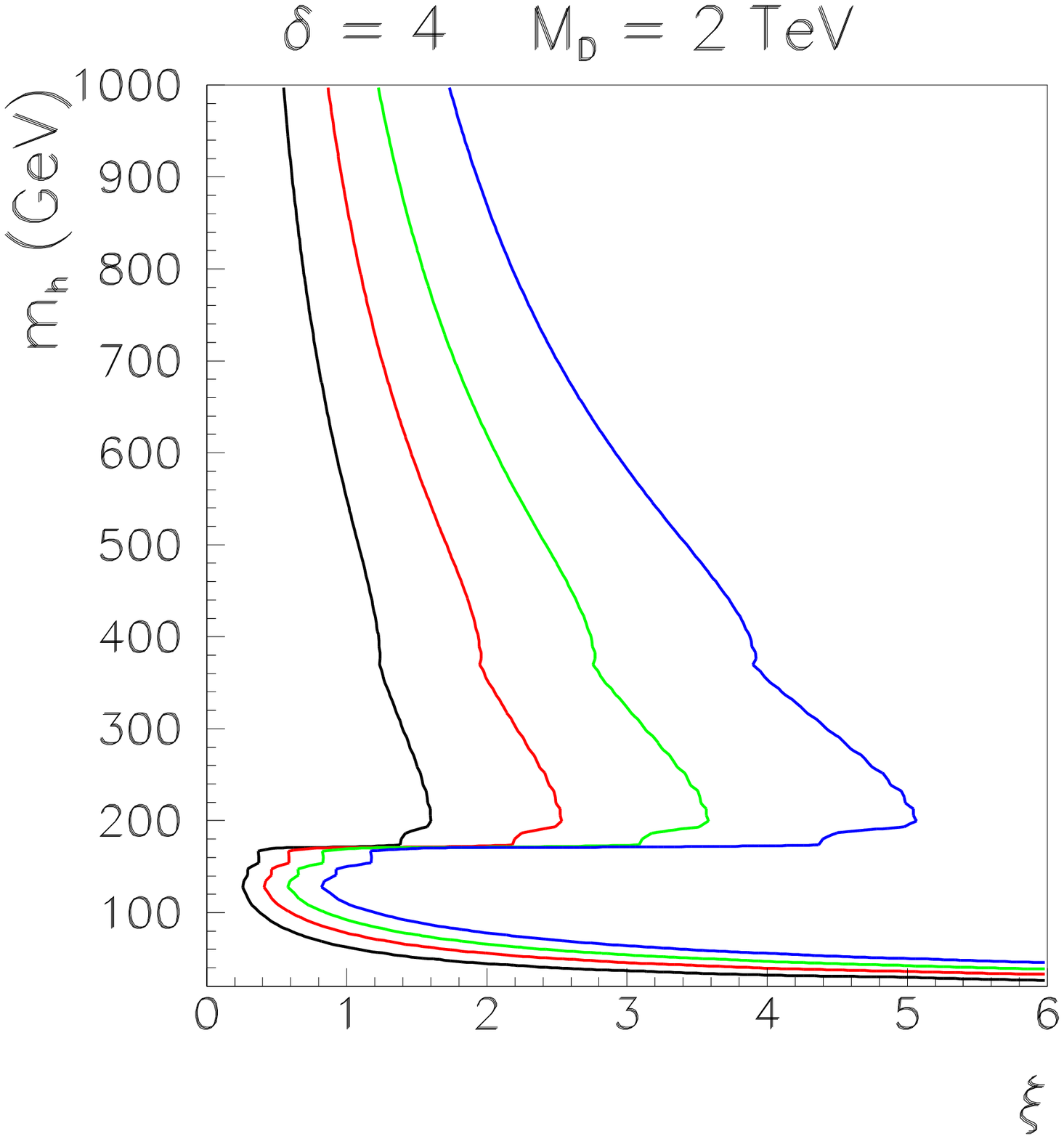}
\caption{We display contours of $R=0.1$, 0.25, 0.5 and 1 (from left to
  right) in the
  $m_h$--$\xi$ plane, for (left plot) $\del=4$ and $\md=1\tev$ and
  (right plot)
  $\del=4$ and $\md=2\tev$.}
\label{Rplot2}
\end{center}
\end{figure}

Of course, the ratio $R$ of Eq.~(\ref{rdef})
can be large even when $(\mh/\md)^4$ is small.  To illustrate the
possibilities for $R$ for typical parameter choices of interest, we
give in Figs.~\ref{Rplot1} and \ref{Rplot2} a few contour plots of $R$
as a function of $\xi$ and $\mh$ for fixed choices of $\md$ and
$\del$.  These plots make it clear that even if $\mh$ is small
compared to $\md$ a substantial invisible width relative to the SM
width is possible for relatively modest values of $\xi$. In
particular, a large value for $R$ is possible for quite small $\xi$
values ({\it i.e.}  $<0.1$) when $\mh$ is below the $WW$ decay
threshold.


\subsection{Diagonal basis derivation of the invisible width}

In this section, we summarize a direct Feynman rule based derivation of
the invisible width, based on first
diagonalizing the Hamiltonian. One begins with a Hamiltonian that includes the widths
of the various states and the mixing between the states. Because of the widths, the
Hamiltonian is intrinsically complex and non-Hermitian.  One must then
diagonalize the Hamiltonian to obtain the mass eigenstates. This is
the procedure that is employed when studying the $K^0$-$\overline K^0$
system. From the diagonalized Hamiltonian one can then derive the
Feynman rules and compute the $WW \to WW$ amplitude using them.

Since the $h$ is the only one of the states with couplings to $WW$ and
$f\anti f$ (that are not suppressed by $1/M_P$) it is the only state
with a substantial inverse propagator imaginary component,
$i\mh\gamh$.  A crucial question is the size of
$\mh\gamh$ relative to $\eps$ and relative to the spacing
between the graviscalar states.  For $\mh\sim 100\gev$ and
$\gamh\sim {\rm few}$~MeV, $\mh\gamh\sim 0.1\gev^2$. The spacing
between graviscalar states depends upon $\del$, ranging from $\sim
{\rm eV}$ for $\del=2$ to $0.1\gev$ for $\del=8$.  The smallest
spacing in mass-squared (taking $\del=2$) is of order $100\gev\times
1~{\rm eV}\sim 10^{-7}~{\gev}^2$, so that it will be important to
understand what happens when the $h$ and $s_i$ states are essentially
degenerate.  The magnitude of $\eps$ is $\sim
(100\gev)^3/10^{19}\gev\sim 10^{-13}\gev^2$, {\it i.e.} much smaller
than either the smallest mass-squared splitting or $\mh\gamh$.
This fact will enter implicitly into some of our expansions.

The relevant mass-squared matrix is (using the Lagrangian form and
defining $\rho\equiv \mh\gamh$)
\beq
\call\propto -\half (\mh^2 -i \rho)h^2-\half \sum_{\vec n>0} m_{\vec n}^2
s_{\vec n}^2+\eps h\sum_{\vec n>0}s_{\vec n}
\label{lagrform}
\eeq
and can be diagonalized to order $\eps^2$ by the following transformations:
\begin{equation}
 h= N \left [ h'+\sum_{\vec m>0}\frac {\eps} {m_h^2-i\rho-m_{\vec m}^2} 
 s_{\vec m}'\right ]
\end{equation}
\beq
 s_n= N_{\vec n}\left [ \sn'- \frac {\eps} {m_h^2-i\rho-m_{\vec n}^2}
  h' -\half {\eps^2\over (m_{\vec n}^2-\mh^2+i\rho)}\sum_{\vec
    m\neq\vec n,\vec n>0,\vec m>0} {1\over m_{\vec m}^2-\mh^2+i\rho}s'_{\vec m}\right]
\eeq
where $h$ and $s_n$ are the original fields before diagonalizing the
Hamiltonian and 
\begin{equation}
N=\left [ 1+\sum_{\vec m>0}\frac {\eps^2} {(m_h^2-i\rho-m_{\vec m}^2)^2} 
\right ]^{-1/2}\,,\quad
\,\,\,\,
N_{\vec n}=\left [1+\frac {\eps^2}{(m_h^2-i\rho-m_{\vec n}^2)^2} \right
]^{-1/2}\,.
\eeq
The corresponding mass squared eigenvalues are
\beq
m_{h'}^2=\mh^2-i\rho+\eps^2\sumn \frac
  1{m_h^2-i\rho-\mn^2}\,,\quad\,\,\,\,\,
m_{\sn'}^2=\mn^2
-\eps^2 \frac {1}  {m_h^2-i\rho-\mn^2}\,.
\eeq

The $\cala(WW\to WW)$ amplitude is then obtained as the sum
 $\cala(WW\to h'\to WW)+\sumn \cala(WW\to  \sn' \to WW)$ and takes the following
 form, as shown in Appendix \ref{appendix}:
\bigskip
\bea
&& \cala_{WW\to h'\to WW}+\sumn\cala_{WW\to  \sn'\to WW}\nn\\
&\sim&\frac i {p^2 -\mh^2+i\rho+\sumn {\eps^2\over \mn^2-\mh^2+i\rho}}
\left(1-\half\sumn\frac {\eps^2}{(\mn^2-\mh^2+i\rho)^2}\right)^2 
+\sumn \frac i {p^2 -\mn^2-{\eps^2\over \mn^2-\mh^2+i\rho}}
\left(\frac {-\eps}{\mn^2-\mh^2+i\rho}\right)^2\nn\\
&\sim&
{\frac i {p^2 -\mh^2+i\rho-\sumn{\eps^2 \over p^2-\mn^2}}\simeq {i \over s -\mh^2+i\mh(\gamh+\gaminv)}\,.
}
\eea
From the final expression we see (again) that the behavior of the
$WW\to WW$ scattering amplitude is indeed that obtained by replacing
the SM Higgs width, $\gamh$, by $\gamh+\gaminv$ in the self energy
portion of the Higgs propagator, with $\gaminv$ as given in
Eq.~(\ref{imsig}). We also stress that this form implies that the
effective cross section for $WW\to WW$ from $s$-channel Higgs
resonance exchange will be suppressed compared to that predicted in
the SM by a factor of $1/(1+R)$. 

Of course, it should be stressed that all of these same remarks apply
to any process where the Higgs is exchanged in the $s$-channel
beginning with a SM state and ending with a SM state.  In any such
channel in which one can scan over the Higgs resonance, the width of
the resonance will be $\gamh+\gaminv$ instead of $\gamh$ and the net
cross section will be suppressed compared to the SM prediction by
$1/(1+R)$.  As studied in \cite{Gunion:1996cn}, there are many
indirect and direct techniques for measuring the total width of the
Higgs resonance.  These range from the very precise determinations by
direct scanning in $\rts$ at a muon collider, which yields excellent
accuracy for the width even for a very narrow Higgs as found for
$\mh\sim 100-150\gev$, to looking at $gg\to h \to ZZ^{(*)}\to 4\ell$ at
a hadron collider, where a direct scan determination of the Higgs
width is possible for $\mh\gsim 200\gev$, \ie\ whenever the width is
larger than a couple of GeV.  Of course, if $\gamh +\gaminv \gg\gamh$,
then the latter will be possible down to considerably lower values of
$\mh$. For example, from Fig.~\ref{paircomp} we see that $\xi\sim 0.6$
will give $\gamh+\gaminv\sim 2\gev$ at $\mh=120\gev$ for $\del=2$ and
$\md=1\tev$.  At a next-generation linear collider, the best technique
for directly scanning the Higgs resonance is to look at the Higgs peak
shape in $\epem\to ZX$ as a function of $\mx$. This again works in the
case of a SM-like width $\gamh$ down to $\mh\sim 200\gev$, extending
to substantially lower values for any case where $\gamh+\gaminv\gsim
2\gev$. In the absence of a muon collider, Ref.~\cite{Gunion:1996cn}
details the means for measuring the effective Higgs width for cases
where $\gamh+\gaminv\lsim 1\gev$ using a subtle
combination of $\gam\gam$ collider data and $\epem$ collider data.

\subsection{Additional contributions to the invisible state production
  coming from direct two-graviscalar production processes}  

In the above, it was assumed that the only way in which invisible
intermediate states contribute to  $WW\to WW$ near the Higgs resonance
is through multiple iterations of $h\to \sn\to
h$ type mixing. However, there are other contributing 
invisible intermediate states as a result of the presence of $WW\to
\sk\sls$ processes.  First, there are $WW\sk\sls$ contact
interactions.  Second, there are the $s$-channel exchange processes involving
$WW \to h+\sn\to \sk\sls$.  (Here, non-$s$-channel diagrams could also
be included, but would yield very small contributions compared to
those we consider, assuming reasonable resolution in the final state.)  To
include the additional resonant sources of invisible state production requires a
significant calculation, performed below. We also show that the
$\sn\sk\sls$ vertices that are present in the theory do not contribute
to the process at hand. Note that we have found it easiest to perform
this calculation in the $h$, $\sn$ basis rather than in the
diagonalized $h'$, $\sn'$ basis. In part, this is because it is the
$\sn$'s that are truly invisible. But also, the $h$ -- $\sn$ basis is
simply easier to use, just as was the case for the mixing width
calculation. At the end, we find that, to a good approximation, the appropriate comparison is
the contribution to the Higgs invisible width coming from $h\to\sn\sm$
decays to that coming from $h$ -- $\sn$ mixing. 

Let us first discuss the $WW\sk\sls$ contact interactions.
These derive from expanding the interaction Lagrangian to order
$\kap^2$ where $\kap=2/\md^{1+\del/2}$. This expansion takes the form
\bea
\int d^4 x \int dy \sqrt{-\what g}\call(\what g)&=& \int d^4 x \int dy
\del(y) \Biggl\{ \left[ \call(\what g)|_{\what g=\eta}-{\kap\over 2}
  h^{\mu\nu}T_{\mu\nu}\right]\nn\\
&&\quad + \kap^2\left[A\call(\what g)|_{\what g=\eta}-B^{\mu\nu}{\del
    \call\over \del \what g^{\mu\nu}}|_{\what g=\eta}\right]\nn\\
&&\quad+\kap^2 h^{\mu\nu}(x,y)\left[\half\int d^4x' {\del^2\call\over
    \del \what g^{\mu\nu}\del\what g^{\rho\sigma}}|_{\what g=\eta}
  h^{\rho\sigma}(x',y)\right]\Biggr\} 
\eea
where 
\beq
A={1\over 8} h^2-{1\over 4}h_{\rho\sigma}h^{\rho\sigma}\,,\quad
B^{\mu\nu}=\half h h^{\mu\nu}-h^{\mu\lambda}h^\nu_\lambda\,,\quad h=h_\mu^\mu\,.
\eeq
After employing the $\del(y)$, which gives a factor of $1/V^\del$,
using the identification ${\kap^2\over V^\del}=1/\mpl^2$, and using
the fact that for an initial $WW$ state we would have ${\del
    \call\over \del \what g^{\mu\nu}}|_{\what g=\eta}\sim \mw^2 W^\mu
  W_\mu+\ldots$,
we find an amplitude contribution to $WW\to \sk\sls$ that is
$\propto \mw^2/\mpl^2$.  Squaring and integrating over a window of
$ds$ of size $\sim \mh \Gamma_{\rm res}$, we get a cross section
contribution of order 
\beq
{\mw^4\over M_D^8}\mh^5\gamres\,.
\eeq
This can be compared to the $s$-channel $h$ exchange contribution
which gives an integrated cross section for $WW\to h\to WW$ of rough
size (assuming that the resolution window size $\Gamma_{\rm res}$ is
substantially larger than $\gamh$)
\beq
g^4\mw^4 {\pi\over \mh\gamh}\,.
\eeq
The ratio of the contact to the $s$-channel contribution is then very
roughly given by
\beq
{\mh^6\gamh\gamres\over \pi g^4 \md^8}\,.
\eeq
This ratio will typically be very small  provided $\gamres$
is of order a few GeV and $\md>1\tev$.

Now let us turn to the cubic interactions that can lead to $WW \to
h+\sn\to \sk\sls$ type 
processes. To do so, we must go to the full $\call$ including
all effects of the mixing term at the cubic level.  A first source of
such cubic interactions comes from the expansion of the
Hilbert-Einstein Lagrangian up to the cubic order in the graviscalar
fields. There are a huge number of terms, but after integrating over
the extra dimensions one finds that all cubic $\sn\sk\sls$ vertices
are proportional to $\del_{\vec n+\vec k+\vec l}$.  This $\delta$
function implies that $|\vec k|+|\vec l|\geq |\vec n|$. Since $m_{\vec
  m}\propto |\vec m|$, this implies that $m_{\vec k}+m_{\vec l}\geq
m_{\vec n}$. However, for the decay $\sn\to\sk+\sls$ to be possible
requires $m_{\vec k}+m_{\vec l}< m_{\vec n}$.  Thus, there is no phase
space available for the decays that could be mediated by these cubic vertices.
  
The $h\sn\sk$ vertices that can mediate the decays of interest derive from the extra
$T_\mu^\mu$ contributions that emerge from $\call_{mix}$ which take
the form (using $H=\rhalf(v+h)$)
\beq
T_\mu^{\mu\,extra}=6\xi \square(H^\dagger H)\,,
\eeq
where 
\beq
\square(H^\dagger H)=(v\square h +h\square h +\partial_\mu h\partial^\mu h)\,.
\eeq
Including this full structure
in the expression for $\call_{int}$, 
gives rise to an interaction term of the form:
\beq
-{\eps\over 6\xi v \mh^2}\sum_{\vec n>0}\sn \left[
4V(h)-\partial^\mu h \partial_\mu h+6\xi (v\square h +h\square h +\partial_\mu h\partial^\mu h)\right]\,,
\eeq
where the full form for $V(h)$ is
\beq
V(h)=\half \mh^2 h^2 +\half {1\over v}\mh^2 h^3+{1\over 8}{1\over v^2}h^4\,.
\eeq
Thus, the full $\call$ takes the form:
\beq
\call =-\half \sumn \sn (\square+\mn^2)\sn -\half h\square h
-\left(1+\sumn {2\over 3}{\eps\over \xi v \mh^2}\sn\right)V(h)
-{\eps\over v\mh^2}\left[v\square h +\partial_\mu h \partial^\mu
h\left(1-{1\over 6\xi}\right)
+h\square h\right]\sumn \sn\,
\eeq
or after partial integration
\bea
\call&=
&
\half\sumn\ptl_\mu\sn\ptl^\mu\sn-\half \sumn\mn^2\sn^2 
+\half \ptl_\mu h\ptl^\mu h
-(1+{2\over 3}{\eps\over \xi v \mh^2}\sumn\sn)V(h)\nn\\
&&
-{\eps\over v\mh^2}\sumn\left(-v\ptl_\mu h\ptl^\mu\sn -\ptl^\mu \sn
  h\ptl_\mu h
-{1\over 6\xi} \sn \partial^\mu h\partial_\mu h\right)\,.
\label{partintlagr}
\eea

At this point, we must now employ the equations of motion
for the $h$ and $\sn$ fields.
For the $\sn$ we have
\beq
(\square+\mn^2) \sn+{\eps\over v\mh^2}(v\square h+\ptl_\mu h\ptl^\mu
h+h\square h)+{2\over 3}{\eps\over \xi v \mh^2}\left( V(h)
-{1\over 4} \partial^\mu h \partial_\mu h\right)=0\,.
\eeq
As regards the $h$, we have:

\bea
&&\square h +{\eps \over v\mh^2}(v+h)\sumn\square\sn
+(1+{2\eps\over 3\xi v\mh^2}\sumn\sn){\ptl V\over\ptl h}
+{\eps\over 3 \xi v m_h^2}\left[ \partial^\mu\left(\sumn\sn \partial_\mu h\right)\right]=0\,.
\label{boxhform}
\eea
Returning to
the original form of $\call$, given in Eq.~(\ref{partintlagr}),
keeping only terms of cubic or quadratic order
in fields, we get
\bea
\call&=& 
\half\sumn\ptl_\mu\sn\ptl^\mu\sn-\half \sumn\mn^2\sn^2 
+\half \ptl_\mu h\ptl^\mu h -\half \mh^2 h^2
-{1\over 2v}\mh^2h^3-{1\over 3}{\eps\over \xi v}\sumn\sn h^2
\nn\\
&&\quad\quad -{\eps\over v\mh^2}(v+h)\square h \sumn\sn -{\eps\over v\mh^2}
\left(1-{1\over 6\xi}\right)\ptl^\mu h\ptl_\mu h \sumn\sn\,.
\label{finallag}
\eea
We now make use of the equations of motion
by substituting the 
 expression for $\square h$ obtained using Eq.~(\ref{boxhform}).  First, we collect the
purely quadratic terms in $\call$ from this substitution.
This means we look for terms linear in the fields
in $\square h$.  To the needed order in $\eps$, we have
\beq
(\square h)_{linear}=-\mh^2 h-\summ{\eps\over \mh^2}\square\sm\,.
\eeq
The  relevant (quadratic) terms arise from the next to last term in
$\call$ in Eq.~(\ref{finallag}) and are
\beq
\call^{\square h}_{quad}=\sumn\eps h\sn+\sumn\summ{\eps^2\over \mh^4}\sn\square\sm\,,
\eeq
the first being our standard mixing term.
Important trilinear terms emerge from the second  term of the next to last term
in Eq.~(\ref{finallag}).
Again keeping only $\calo(\eps^2)$ or lower we have:
\beq
\call^{\square h}_{cubic}=-{\eps \over v \mh^2}h\sumn\sn(\square h)_{linear}-\sumn{\eps\over \mh^2}\sn (\square h)_{quad}\,.
\eeq
To the order needed,
\beq
(\square h)_{quad}=-{3\over 2}{\mh^2\over v}h^2-\summ{2\eps\over 3\xi
  v }\sm h-\summ{\eps \over v\mh^2}\square\sm h-{\eps\over 3v
  m_h^2\xi}\sumn\left(\partial^\mu \sn \partial_\mu h +\sn \square h\right)
\eeq
yielding
\beq
\call^{\square h}_{cubic}=\sumn\left[{5\over 2}
{\eps\over v}\sn h^2+{1\over 3}{\eps^2\over \xi v\mh^2}h\sn \sum_{\vec
  m>0}\sm+2{\eps^2 \over v\mh^4}\sn h \sum_{\vec m>0}\square\sm
+{\eps^2\over 3 v m_h^4 \xi}\sn \summ\partial^\mu \sm \partial_\mu h
\right]\,.
\eeq
So, putting it all together, we find the following trilinear Lagrangian
\bea
\call_{cubic}&=&-{1\over 2v}\mh^2 h^3 -{\eps\over 3\xi v}h^2\sumn
\sn-{\eps\over v\mh^2}\left(1-{1\over 6\xi}\right)\ptl^\mu h\ptl_\mu h\sumn\sn
+\call_{cubic}^{\square h}\nn\\
&=&-{1\over 2v}\mh^2 h^3 -{\eps\over v\mh^2}\left(1-{1\over 6 \xi}\right)\ptl^\mu h\ptl_\mu
h\sumn\sn+{\eps\over 2v}\left(5-{2\over 3\xi }\right)h^2\sumn\sn\nn \\
&&\quad
 +{1\over 3}{\eps^2\over \xi v\mh^2}h\sumn\sn \sum_{\vec m>0}\sm+2{\eps^2 \over v\mh^4}h\sumn\sn  \sum_{\vec m>0}\square\sm
+{\eps^2\over 3v m_h^4 \xi}\sumn \sn \summ \partial^\mu
\sm \partial_\mu h 
\eea
of which it is the latter two terms that give the vertices of interest.
The effective cubic Lagrangian for our purposes then becomes (after
using $\square s_{\vec m}\to -m_{\vec m}^2 s_{\vec m}$ and relabeling indices)
\beq
\call_{cubic}=\eps^2\sumk\suml\biggl[{1\over 2\xi v\mh^2}h\sls \sk -{2
  \over v\mh^4}\left(1-{1\over 6\xi}\right)h\sls  \mk^2\sk\biggr]\,.
\eeq
The $h\sk\sls$ vertex (accounting for the many different Wick's
contractions) takes the form $\eps^2 \gkl$ where
\bea
\gkl&=&{1\over \xi v\mh^2}-{2\over
v\mh^4}\left(1-{1\over 6 \xi}\right)(\mk^2+\ml^2) \,.
\eea

The invisible $h$ width arising from these interactions takes the form
\beq
\Gamma(h\to graviscalar~pairs)=\half \sum_{\vec k>0,\vec l>0}
{1\over 16\pi m_h^3}|\eps^2 g_{\vec l,\vec k}|^2\lam(m_h^2,m_{\vec
  k}^2,m_{\vec l}^2)\,,
\label{gampair}
\eeq
where
$\lam(a,b,c)=\left[a^2+b^2+c^2-2ab-2ac-2bc\right]^{1/2}$
is the usual two-body phase space factor and the
$\half$ is required to avoid double counting states.
To compute this width numerically, we employ Eq.~(\ref{gampair}) and write  
\bea
\half\sum_{\vec k>0,\vec l>0}&=&\half \left(\half\right)^2 \int dm_k^2
dm_l^2
\rhodel(m_k^2)\rhodel(m_l^2)\nn\\
&=&\half\left(\half\right)^2\rhodel^2(\mh^2)\mh^4\int
dx (\sqrt x)^{\del-2}\int dy  (\sqrt y)^{\del -2}\nn\\
&=&
\half\left(\half\right)^2\rhodel^2(\mh^2) \mh^4\int_0^1 dx x^{{\del\over
    2}-1}\int_0^1 dz 
(1-\sqrt x)^2 \left[(1-\sqrt x)^2z\right]^{{\del\over 2}-1}
\eea
where we have defined 
\beq
x\equiv {m_l^2\over \mh^2}\,,\quad y\equiv (1-\sqrt x)^2 z \equiv
{m_k^2\over \mh^2}
\eeq
and used the definition of the density, $\rhodel$, given in Eq.~(\ref{rhodeldef}).
The integration limits derive from the presence of the $\lam$
kinematic phase space factor, which reduces to
\beq
\lam(m_h^2,m_{\vec k}^2,m_{\vec l}^2)=\mh^2 \lam(1,x,y)=\mh^2(1-\sqrt x)\sqrt{1-z}\left[1+x(1-z)+2\sqrt x(1+z)-z\right]^{1/2}\,,
\eeq
from which one immediately sees that phase space runs out at $x=1$ or $z=1$.
In terms of the $x$ and $z$ variables, we have
\bea
\gkl&=&{1\over v\mh^2}\Biggl\{{1\over \xi}-2\left(1-{1\over 6\xi}\right)(x+(1-\sqrt x)^2 z)\Biggr\}\,.
\eea

The final
expression for $\Gamma(h\to graviscalar~pairs)$ can be written
in terms of the integral
\bea
I&=&{1\over 4}\int_0^1 dx\int_0^1 dz (1-\sqrt x)^{\del+1}\sqrt{1-z}
\left[1+x(1-z)+2\sqrt x(1+z)-z\right]^{1/2} x^{\del/2-1}z^{\del/2-1}\nn\\
&&\times \Biggl|{1\over \xi}-2\left(1-{1\over 6\xi}\right)(x+(1-\sqrt x)^2z)\Biggr|^2\,.
\eea
$I$ behaves as $1/\xi^2$ at small $\xi$, reaches a
minimum near $\xi=1.5$ due to the cancellations implicit in $\gkl$, and
ultimately asymptotes (quite slowly) to a constant value of $I\to
0.011$ for $\del=2$ ($I\to 0.00024$ for $\del=4$) at $\xi\to \infty$.
We plot $\xi^2\,I$ as a function of $\xi$ for the
$\del=2$ and $\del=4$ cases in Fig.~\ref{intplot}.
Clearly, $I$ decreases rapidly as $\del$ increases. 
As a result,  $\Gamma(h\to graviscalar~pairs)$ is only
significant compared to $\Gamma(h\to graviscalar)$
if $\del\leq 4$.
\begin{figure}[htbp]
\begin{center}
\includegraphics[width=8.0cm,height=5.0cm]{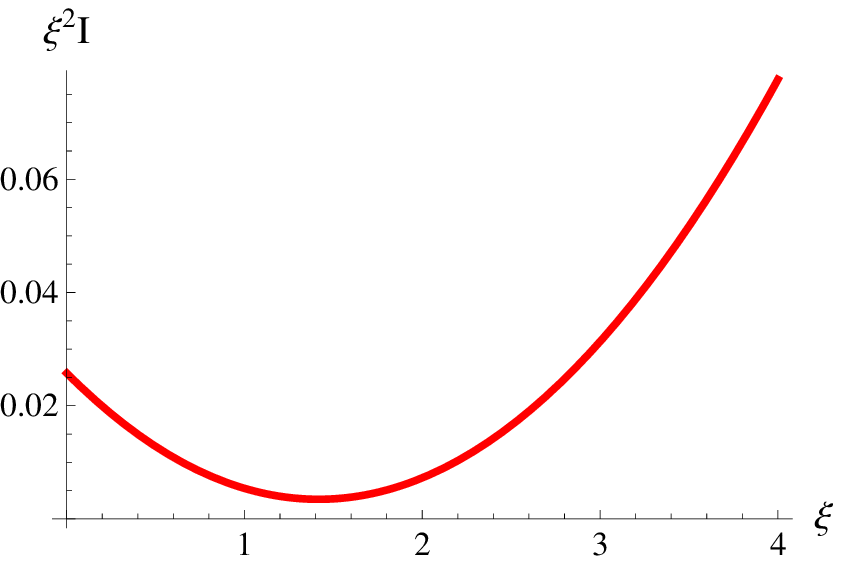} 
\includegraphics[width=8.0cm,height=5.0cm]{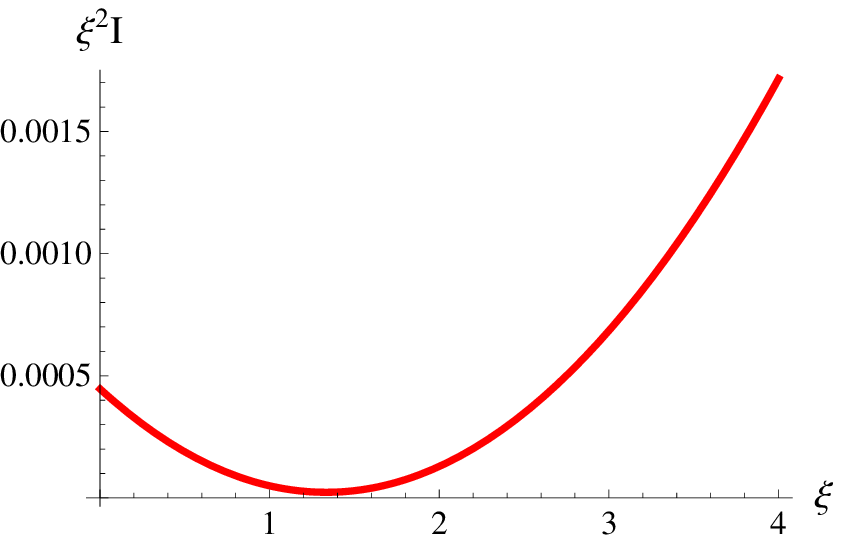} 
\caption{We plot $\xi^2\, I$ as a function of $\xi$ for the cases of
  $\del=2$ and $\del=4$.}
\label{intplot}
\end{center}
\end{figure}

In terms of $I$, we find
\bea
\Gamma(h\to graviscalar~pairs)&=&{1\over 32\pi}{m_h^{2\del-5}
\eps^4\over v^2} {M_P^4\over \md^{4+2\del}}
\left[{\pi^{\del/2}\over \Gamma(\del/2)}\right]^2 I\nn\\
&=& {18\over \pi} {m_h^{3+2\del}v^2\over \md^{4+2\del}}\xi^4\left({\del-1\over\del+2}\right)^2 \left[{\pi^{\del/2}\over \Gamma(\del/2)}\right]^2 I\,.
\eea
This is to be compared to the direct mixing invisible width, which
in terms of $\eps$ takes the form:
\beq
\Gamma(h\to graviscalar)= {\pi\over 2} m_h^{\del -3}\eps^2 {M_P^2\over \md^{2+\del}}
{\pi^{\del/2}\over \Gamma(\del/2)}\,.
\eeq
We obtain
\bea
{\Gamma(h\to graviscalar~pairs)\over \Gamma(h\to graviscalar)}&=&
{1\over 16\pi^2}{m_h^{\del -2}\eps ^2\over v^2} {M_P^2\over \md^{2+\del}}{\pi^{\del/2}\over \Gamma(\del/2)}I\nn\\
&=& {3(\del -1)\over 2\pi^2(\del+2)}\xi^2 \left({m_h\over \md}\right)^{2+\del}{\pi^{\del/2}\over \Gamma(\del/2)}I\,.
\label{ratioform}
\eea
From this result, we immediately see that unless $m_h$ is
comparable to or larger than $\md$ the pair invisible width will
be much smaller than the mixing invisible width unless $\xi^2$ is
large enough, $\xi\sim 10$, to overcome the numerically small value of
$I\sim 0.011$ at large $\xi$. However, for such large $\xi$ the mixing
invisible width is typically huge. At small $\xi$, since $\xi^2 I$ approaches a
constant value so does $\Gamma(h\to graviscalar~pairs)/ \Gamma(h\to
  graviscalar)$.
 
\begin{figure}[htbp]
\begin{center}
\includegraphics[width=8.0cm,height=5.0cm]{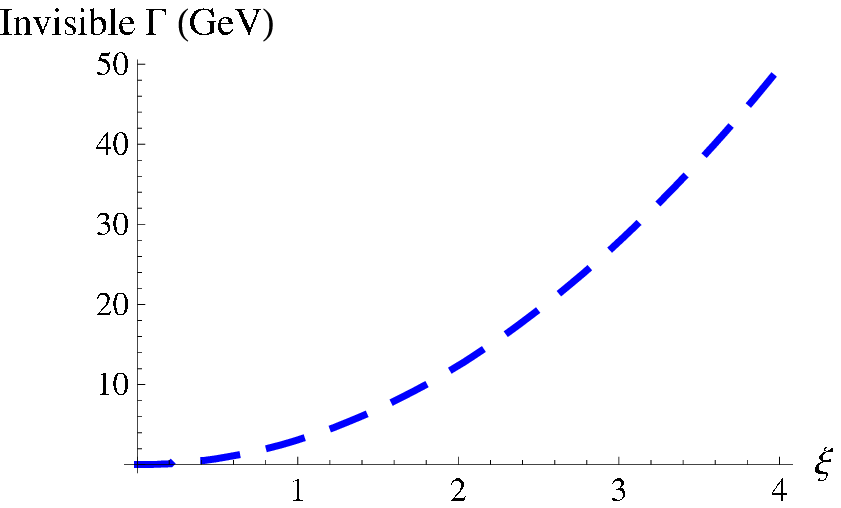} 
\includegraphics[width=8.0cm,height=5.0cm]{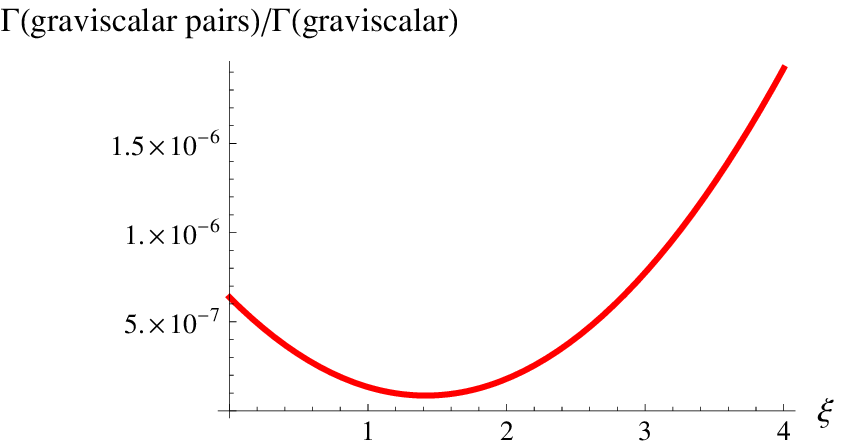}
\caption{We display $\Gamma_{inv}=\Gamma(h\to graviscalar)$ and $\Gamma(h\to graviscalar~pairs)/\Gamma(h\to  graviscalar)$ for 
  $\del=2$, $\md=1\tev$ and $\mh=120\gev$.}
\label{paircomp}
\end{center}
\end{figure}

\begin{figure}[h!]
\begin{center}
\includegraphics[width=8.0cm,height=5.0cm]{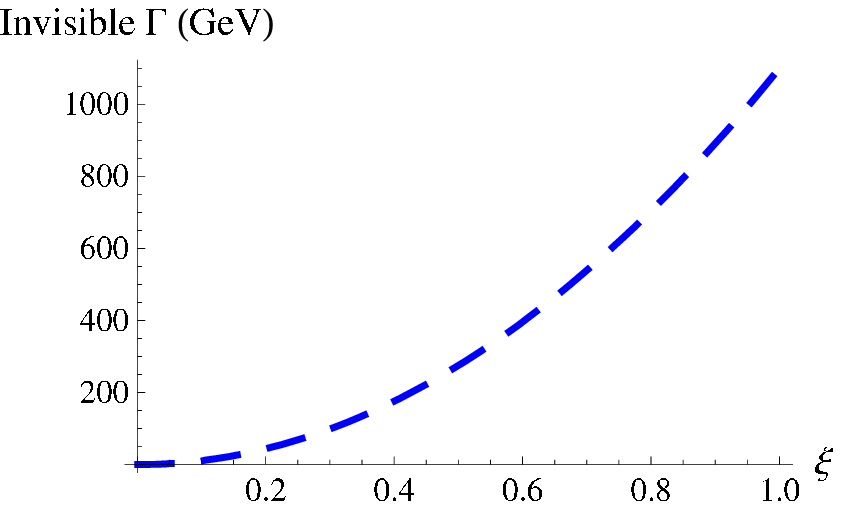} 
\includegraphics[width=8.0cm,height=5.0cm]{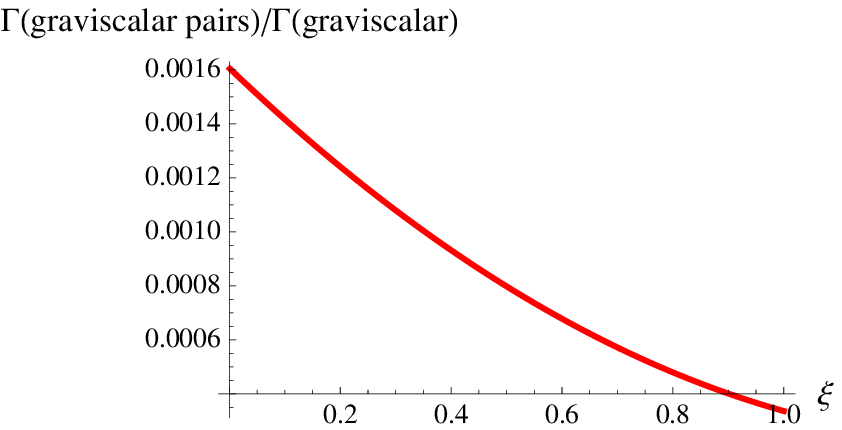}
\caption{We display $\Gamma_{inv}=\Gamma(h\to graviscalar)$ and $\Gamma(h\to graviscalar~pairs)/\Gamma(h\to  graviscalar)$ for 
  $\del=2$, $\md=1\tev$ and $\mh=850\gev$.}
\label{paircomp2}
\end{center}
\end{figure}

\begin{figure}[h!]
\begin{center}
\includegraphics[width=8.0cm,height=5.0cm]{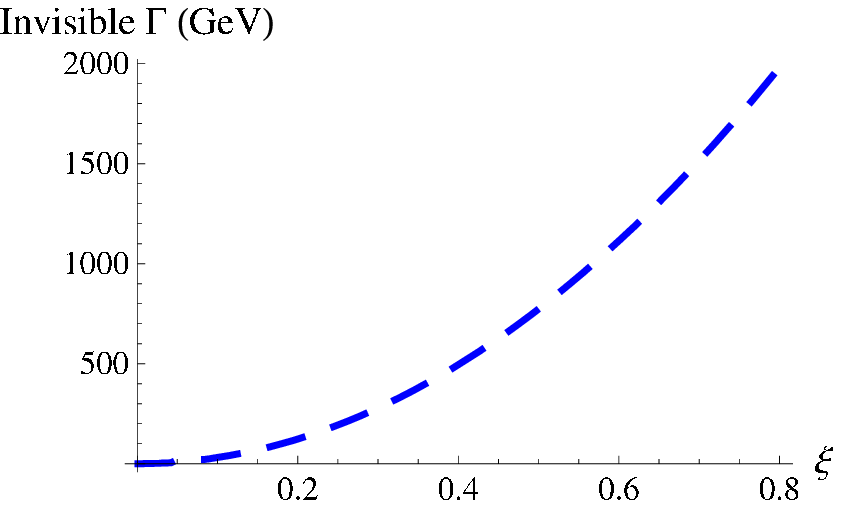} 
\includegraphics[width=8.0cm,height=5.0cm]{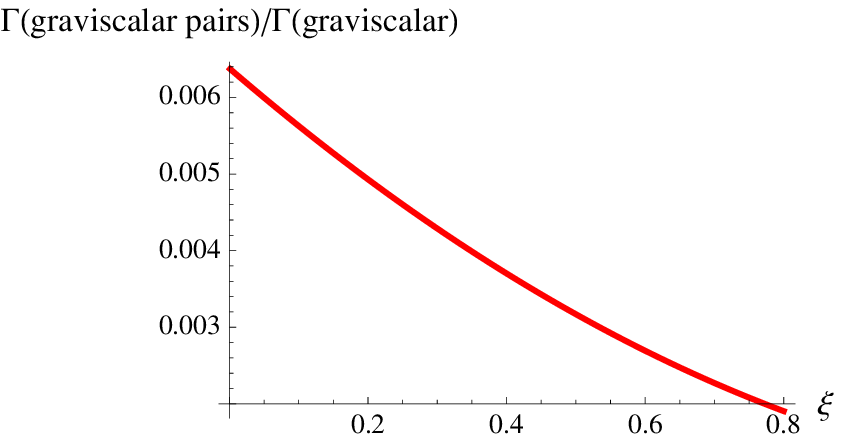}
\caption{We display $\Gamma_{inv}=\Gamma(h\to graviscalar)$ and $\Gamma(h\to graviscalar~pairs)/\Gamma(h\to  graviscalar)$ for 
  $\del=2$, $\md=1\tev$ and $\mh=1200\gev$.}
\label{paircomp3}
\end{center}
\end{figure}

\begin{figure}[h!]
\begin{center}
\includegraphics[width=8.0cm,height=5.0cm]{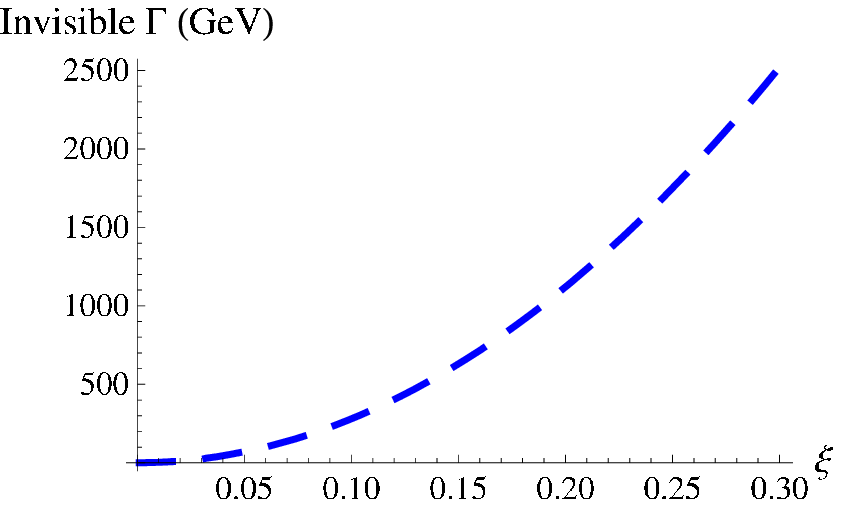} 
\includegraphics[width=8.0cm,height=5.0cm]{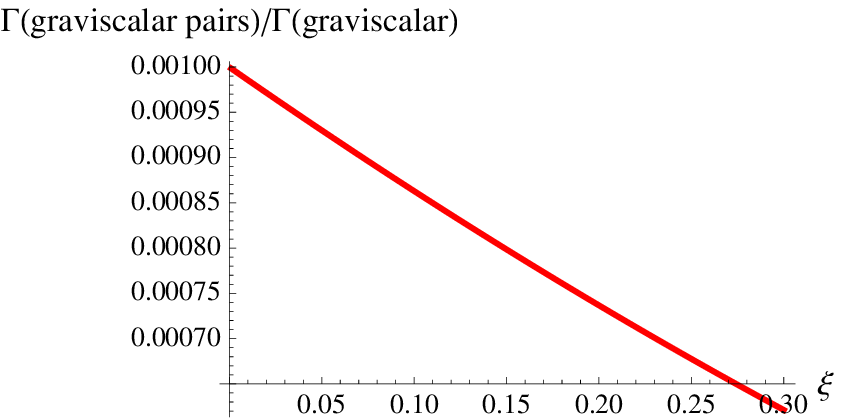}
\caption{We display $\Gamma_{inv}=\Gamma(h\to graviscalar)$ and $\Gamma(h\to graviscalar~pairs)/\Gamma(h\to  graviscalar)$ for 
  $\del=4$, $\md=1\tev$ and $\mh=1200\gev$.}
\label{paircomp4}
\end{center}
\end{figure}

To illustrate, we display in Figs.~\ref{paircomp}, \ref{paircomp2} and
\ref{paircomp3} the mixing invisible width $\Gamma(h\to graviscalar)$
and  the ratio $\Gamma(h\to 
graviscalar~pairs)/\Gamma(h\to  graviscalar)$ as a function of $\xi$
for the cases of $\del=2$, $\md=1\tev$ (the approximate lower limit for $\del=2$ from
Tevatron data) and a selection of Higgs masses: $\mh=120\gev$,
$\mh=850\gev$ and $\mh=1200\gev$. One finds that the graviscalar-pair
to graviscalar-mixing width ratio is very small for the light Higgs case and 
only as large as a percent even for $\mh$ somewhat above $1\tev$, a
range of $\mh$ that becomes questionable from the point of view of unitarity 
for $WW\to WW$ scattering. For $\del=4$, the ratio is even smaller as
apparent from the example of $\del=4$, $\md=1\tev$ and $\mh=1200\gev$
plotted in Fig.~\ref{paircomp4}.  In the above plots, one should presumably
not take seriously the regions at larger $\xi$ in the $\mh=850\gev$
and $1200\gev$ cases for which $\gaminv$ becomes substantially larger
than $\mh$.

\section*{CONCLUSIONS}

We computed the amplitude for a process such as $WW\to WW$ in the
presence of Higgs-graviscalar mixing.  Using a direct Feynman diagram
approach we performed the calculation in two different bases: (a) the
non-mass-diagonal (Lagrangian) basis and (b) the diagonalized mass
basis. Of course, identical results were obtained but the comparison
provides a few pedagogical and intuitive insights.  Ignoring small
corrections from wave-function renormalization, the resulting
amplitude is equivalent to that for exchange of a single effective
Higgs state with SM coupling to $WW$ but effective width given by
$\gamh+\gaminv$, where $\gamh$ is the Higgs width as computed in the
standard model and (up to the factor of 2 correction that we found)
$\gaminv$ is the width obtained via the original technique of
Ref.~\cite{Giudice:2000av}. In particular, one would not observe a sum
of Breit-Wigners, one of width $\gamh$ overlaying a superposition of
many much narrower resonances. Integration over the full resonance
will yield a $WW\to WW$ cross section that is suppressed compared to
the SM result by the factor $1/(1+R)$, where $R=\gaminv/\gamh$. Of
course, the width and total cross section for any process mediated by
Higgs exchange will be affected in exactly the same way as is $WW\to
WW$. For example, for any choices of $\del$, $\xi$ and
  $\md$ such that $\gamh+\gaminv\gsim 2\gev$ it will be
possible to directly measure this net width in the process $gg\to
Higgs\to ZZ\to 4\ell$ by looking at the peak shape in $M_{4\ell}$. 

Using the Feynman diagram technique we were also able to perform a
direct computation of the invisible width from $WW\to h\to \sum_{\vec
  n,\vec k} \sn\sk$ on resonance pair production of graviscalars.  We
found that this width was typically much smaller than the invisible
width from Higgs-graviscalar mixing unless $\mh>\md$ ($\md$ being the
effective $4+\delta$ dimensional Planck scale). However, the $\mh>\md$
region of parameter space (a) is problematical from the point of view
of $WW\to WW$ unitarity given that $\md>1\tev$ and (b) is such that
many other corrections to the invisible width are possibly
present. Nonetheless, for $\mh>\md$ including the pair width would be
necessary for precision comparison between theory and experiment.

\vskip1cm
\noindent

\section*{ACKNOWLEDGMENTS}
D. Dominici was partially supported by MIUR under the contract
PRIN-2006020509.
JFG is supported by the U.S. Department of Energy
under grant No. DE-FG03-91ER40674.
JFG would also like to thank the University of Florence, the
INFN, the Aspen Center for Physics and the Kavli Institute for
Theoretical Physics for support  
during various parts of this project.

\appendix
\section{Feynman diagram derivation of the invisible width and full
$WW\to WW$ scattering amplitude form}
\label{appendix}

  To
begin with, it is useful to understand how the calculation should be
done using the example of a small number of states.  We illustrate
using 3 states.  The three states are the $h$ (Higgs) with
mass-squared $\mu^2$, a graviscalar $s_1$ with mass $m_1$ and a second
graviscalar $s_2$ with mass $m_2$.

Since the $h$ is the only one of the states with couplings to 
$WW$ and $f\anti f$ (that are not suppressed by $1/M_P$) it is the only
state with a substantial imaginary component, $i\half\mu\Gamma_h$.

The relevant mass-squared matrix is (using Eq.~(\ref{lagrform})) 
\begin{equation}
\half M^2 \equiv {1\over 2}\left(\begin{array}{ccc} \mu^2-i\mu\Gamma_h & -\epsilon &-\epsilon \\
-\epsilon& m_1^2 & 0 \\
-\epsilon & 0 & m_2^2\\
\end{array}
\right)\,.
\end{equation}
This $3\times 3$ matrix is 
a complex symmetric matrix, that is not hermitian.  
It can be diagonalized by an orthogonal transformation
(given that the eigenvalues are distinct). To sufficient order, the eigenvalues are 
\beq
\mu^2-i\rho+\eps^2{1\over\mu^2-i\rho-m_1^2}+\eps^2{1\over\mu^2-i\rho-m_2^2}\,,\quad m_1^2+{\eps^2\over m_1^2-\mu^2+i\rho}\,,\quad m_2^2+{\eps^2\over m_2^2-\mu^2+i\rho}\,,
\label{eig}
\eeq
where we have written $\rho\equiv \mu\Gamma_h$.
The eigenvectors of the $3\times 3$ matrix are (dropping terms
of ${\cal O}(\eps^3)$ and higher)
\bea
w_1&=&\left(\begin{array}{c} 1 -\frac 1 2 \eps^2 (\frac {1} {(m_1^2-\mu^2+i\rho)^2}+
\frac {1}{(m_2^2-\mu^2+i\rho)^2})\\
{-\eps\over \mu^2-m_1^2-i\rho}\\
{-\eps\over \mu^2-m_2^2-i\rho} \\
\end{array}
\right)\nonumber\\
w_2&=&\left(\begin{array}{c} {\eps\over \mu^2-i\rho-m_1^2} \\
1-\half \frac {\eps^2}{(m_1^2-\mu^2+i\rho)^2}\\
-\half\frac {\eps^2}{ ( m_1^2-\mu^2-i\rho)(m_2^2-\mu^2-i\rho)} \\
\end{array}
\right)\nonumber\\
w_3&=& 
\left(\begin{array}{c} {\eps\over \mu^2-i\rho-m_2^2} \\
-\half\frac {\eps^2}{ ( m_2^2-\mu^2-i\rho)(m_1^2-\mu^2-i\rho)} \\
1-\half \frac {\eps^2}{(m_2^2-\mu^2+i\rho)^2}\\
\end{array}
\right)\,.
\eea
These form a normalized orthogonal basis 
in the sense that $w_i^T w_i=\delta_{ij}$ for $i,j=1,2,3$
and not
$w_i^\dagger w_j=\delta_{ij}$. 
The matrix which diagonalizes the mass-squared matrix is built as
\beq
T=\{w_1,w_2,w_3\}
\eeq
and one can check  that (to order $\eps^2$)
\beq
T^{-1}M^2 T=M^2_D
\eeq
where $M^2_D$ is the diagonal matrix containing the eigenvalues
of Eq. (\ref{eig}). Also, to order $\eps^2$,
$T^{-1}$ coincides with the transpose of $T$.
Therefore the relation between $\phi=\{h,s_1,s_2\}$ and the mass eigenstates
$\phi'=\{h',s_1',s_2'\}$  is given by
\beq
\phi'=T^T \phi~~~~~~~~~\phi=T \phi'.
\eeq
from which we obtain
\bea
h&=&w_1(1) h' +w_2(1)s_1' +w_3(1) s_2'\nn\\
&=&\left[1 -\half \eps^2 \left(\frac {1}{(m_1^2-\mu^2+i\rho)^2}+
\frac {1}{(m_2^2-\mu^2+i\rho)^2}\right)\right] h' 
 +{\eps\over \mu^2-i\rho-m_1^2}s_1' +{\eps\over \mu^2-i\rho-m_2^2}s_2'
\label{hdef}
\eea
where $w_i(1)$ is the first component of the vector $w_i$.
Similarly, we have 
\bea
s_1&=&w_1(2) h' +w_2(2)s_1' +w_3(2) s_2'\nn\\
&\simeq & {-\eps\over \mu^2-m_1^2-i\rho} h'+\left(1-\half \frac
  {\eps^2}{(m_1^2-\mu^2+i\rho)^2}\right)s_1'
-\half\frac {\eps^2}{ ( m_2^2-\mu^2-i\rho)(m_1^2-\mu^2-i\rho)}s_2'\,,
\label{sdef}
\eea
and similarly for $s_2$.

Of course, since the transformation is orthogonal the kinetic terms
for the original $h$, $s_1$ and $s_2$ states transform into
\beq
\half\left(\partial_\mu h'\partial^\mu h' +\partial_\mu s_1'\partial^\mu s_1'+
\partial_\mu s_2'\partial^\mu s_2'\right)
\eeq
where we recall that the fields $h,s_1,s_2$ were
real while these new $h',s_1',s_2'$ now have (small) complex
components.

For later use, we will want the $WW$ coupling of each of the 3 $h',s_1',s_2'$ states.
This comes entirely from the $h$ part of each state giving (relative
to the SM coupling $g_{WWh}$)
\bea
g_{WWh'}&=&1 -\frac 1 2 \eps^2 \left[
\frac {1}{(m_1^2-\mu^2+i\rho)^2}+\frac
{1}{(m_2^2-\mu^2+i\rho)^2}\right]\nn\\
g_{WWs_1'}&=&{\eps\over \mu^2-i\rho-m_1^2}\nn\\
g_{WWs_2'}&=&{\eps\over \mu^2-i\rho-m_2^2}\,.
\eea
These are converted to Feynman rules for the vertices
as usual.  Note, that the couplings are complex; this
will be important in what follows. 

From the above, the generalization to many $s_i$ states is apparent.
For the $h'$ mass-squared we find
\bea
m_{h'}^2&=&\mu^2-i\rho+\eps^2\sumn \frac
  1{\mu^2-i\rho-\mn^2} \nn\\
&=& \mu^2-i\rho +\eps^2\sumn {\mu^2-\mn^2+i\rho\over
  (\mu^2-\mn^2)^2+\rho^2}\nn\\
&\simeq&\mu^2\left(1+{\eps^2\over \mu^2}\Re\left[\sumn \frac
1{\mu^2-i\rho-\mn^2}\right]\right)-i\left(\rho-\eps^2\sumn
\pi\del(\mu^2-\mn^2)\right)\nn\\
&\sim &\mu^2-i(\rho-\rho_{inv})
\eea
where we have neglected the $\Re$ term (which is of order
$\mu^4/\md^4$).  Note that in the $h'$ propagator, $i/(p^2-\mhprime^2)$,
the invisible width comes in with what appears to be a ``wrong''
sign. However, as we will see in the following,
in a typical physical amplitude, one must sum over both
$h'$ and $\sn'$ exchanges.  The sum produces an effective propagator
of form $i/[p^2-\mu^2+i(\rho+\rhoinv)]$.  Below we show this
explicitly for $WW\to WW$ scattering.   

First, however, we must give expressions for the  $WW$ couplings to
the $\hprime$ and $\sn'$ states. We have 
\beq
g_{WWh'}=1 -\frac 1 2 \sumn \eps^2 
\frac {1}{(\mn^2-\mu^2+i\rho)^2}
\,,\quad 
g_{WWs'_{\vec n}}={\eps\over \mu^2-i\rho-\mn^2}\,.
\eeq
The $WW\to WW$ amplitude is then the sum
 $WW\to h'\to WW+\sumn WW\to  \sn' \to WW$ and takes the form:
\vskip .5in
\bea
&& \cala_{WW\to h'\to WW}+\sumn\cala_{WW\to  \sn'\to WW}\nn\\
&\sim&\frac i {p^2 -\mu^2+i\rho+\sumn {\eps^2\over \mn^2-\mu^2+i\rho}}
\left(1-\half\sumn\frac {\eps^2}{(\mn^2-\mu^2+i\rho)^2}\right)^2 
+\sumn \frac i {p^2 -\mn^2-{\eps^2\over \mn^2-\mu^2+i\rho}}
\left(\frac {-\eps}{\mn^2-\mu^2+i\rho}\right)^2\nn\\
&\sim &
\frac i {p^2 -\mu^2+i\rho+\sumn{\eps^2\over \mn^2-\mu^2+i\rho}}
\left(1-\sumn \frac {\eps^2}{(\mn^2-\mu^2+i\rho)^2}\right) 
+\sumn\frac i {p^2 -\mn^2-{\eps^2\over \mn^2-\mu^2+i\rho}}
\frac {\eps^2}{\left(\mn^2-\mu^2+i\rho\right)^2}\nn\\
&\sim&
\frac i {p^2 -\mu^2+i\rho+\sumn{\eps^2\over \mn^2-\mu^2+i\rho}}
\left[1-\sumn\frac {\eps^2}{(\mn^2-\mu^2+i\rho)^2}
+\sumn\frac{p^2-\mu^2+i\rho }
{p^2-\mn^2}
\frac {\eps^2}{\left(\mn^2-\mu^2+i\rho\right)^2}\right]\nn\\
&\sim&
\frac i {p^2 -\mu^2+i\rho+\sumn {\eps^2\over \mn^2-\mu^2+i\rho}}
\left[1+\sumn\frac {\eps^2}{(\mn^2-\mu^2+i\rho)^2}
\frac{\left((p^2-\mu^2+i\rho)-(p^2-\mn^2)\right) }
{p^2-\mn^2}
\right]\nn\\
&\sim&
\frac i {p^2 -\mu^2+i\rho+\sumn {\eps^2\over \mn^2-\mu^2+i\rho}}
\left[1+\sumn\frac {\eps^2}{(\mn^2-\mu^2+i\rho)(p^2-\mn^2)}
\right]\nn\\
&\sim&
\frac i {\left[p^2 -\mu^2+i\rho+\sumn {\eps^2\over \mn^2-\mu^2+i\rho}\right]
\left[1-\sumn\frac {\eps^2}{(\mn^2-\mu^2+i\rho)(p^2-\mn^2)
}\right]}\nn\\
&\sim&
\frac i {p^2 -\mu^2+i\rho+\eps^2\sumn\left(
{1\over \mn^2-\mu^2+i\rho}
-\frac {p^2-\mu^2+i\rho}{(\mn^2-\mu^2+i\rho)(p^2-\mn^2)
}\right)}\nn\\
&\sim&
\frac i {p^2 -\mu^2+i\rho+\eps^2\sumn\left(
{1\over \mn^2-\mu^2+i\rho}
-{1\over p^2-\mn^2}-{1\over \mn^2-\mu^2+i\rho}\right)}\nn\\
&\sim&
\frac i {p^2 -\mu^2+i\rho-\sumn{\eps^2 \over p^2-\mn^2}
}\nn\\
&\sim&
\frac i {p^2 -\mu^2+i\rho-\half \int dm^2\rho_\delta (m^2){\eps^2 \over p^2-m^2+i\eps'}
}\nn\\
&\sim&
\frac i {p^2 -\mu^2+i\rho+F(p^2)+iG(p^2)
}\,,
\label{stepequation}
\eea
where 
\beq F(p^2)=-\eps^2 P
\half \left[\int {dm^2\rhodel(m^2)\over p^2-m^2}\right],\quad G(p^2)=\half\pi\eps^2\rhodel(p^2)\,.
\eeq 
The most critical step in the above is the transition from line 6 to
line 7 of Eq.~(\ref{stepequation}) in which we presume the higher order terms of order $\eps^4$
and so forth organize into the correct geometric series (as they did
in the approach of Ref.~\cite{Giudice:2000av}).
We next write $F(p^2)=F(\mh^2)+
(p^2-\mh^2)F'(\mh^2)+\ldots$, 
where $\mh^2-\mu^2+F(\mh^2)=0$ and drop
the $\ldots$. We also approximate $G(p^2)=G(\mh^2)$ and use the result
of Eq.~(\ref{gaminvdef}) to obtain
\beq
\cala_{WW\to WW}\sim{i\over (p^2-\mh^2)[1+F'(\mh^2)]+i\mh(\Gamma_h+\Gamma_{inv})}\,. 
\eeq  
The result above shows that the $WW\to WW$ scattering amplitude is
indeed equivalent to that for a single Higgs exchange with total width
given by $\gamh+\gaminv$, aside  from wave-function renormalization
associated with the graviscalar
mixing. However, the wave-function renormalization correction is small
since $F'\sim \calo(\mh^4/\md^4)$ is very small for  
$\mh\ll \md$, as required for the model to be fully trustworthy.

If we now take the absolute
square of this form and integrate over $p^2$,
we obtain
\bea
\int dp^2 \left|{i\over (p^2-\mh^2)[1+F'(\mh^2)]+i\mu(\Gamma_h+\Gamma_{inv})}\right|^2
&=&\int dp^2 {1\over (p^2-\mh^2)^2[1+F']^2+\mh^2(\Gamma_h+\Gamma_{inv})^2}\nn\\
&\sim & {1\over 1+F'}{\pi \over \mh(\Gamma_h+\Gamma_{inv})}\,,
\label{dsint}
\eea
which is to be compared to the result we would have obtained
in the absence of graviscalars, which is  $\sim {\pi \over \mh \Gamma_h}$.
Eq.~(\ref{dsint}) shows that (neglecting wave-function
renormalization) the integral over $p^2=s_{WW}$ of
the $WW\to WW$ scattering amplitude gives the $WW$ partial width
(which has been implicitly set to unity for this discussion)
divided by the total width including the graviscalar mixing
contribution.

The above discussion neglects a small correction to the $h'$ and $\sn'$
couplings deriving from the $\sn$ couplings.  These take the form
(in the general case):
\beq
g_{WW\sn}={\eps(1-6\xi)\mw^2 \over 3\xi v
  \mu^2}=g_{WWh}{\eps(1-6\xi)\mw \over 3\xi gv \mu^2}=g_{WWh}{\eps(1-6
\xi)  \over 6\xi  \mu^2}\,,
\eeq
where we were careful to rewrite $T_\mu^\mu=-m_V^2 \vec V_\mu \cdot
\vec V^\mu=-m_V^2(2W_\mu^+W^{\mu\,-}+W^3_\mu W^{3\,\mu})$, giving rise
to an extra factor of $2$ for $W^+W^-$ couplings. This result actually
applies to all types of couplings: in particular, we also have
\beq
g_{f\anti f \sn}=-{\eps(1-6\xi) m_f \over 6\xi v \mu^2}=g_{f\anti f
  h}{\eps(1-6\xi)\over 6\xi \mu^2}\,. 
\eeq
We define
the common $WW$ and $f\anti f$ ratio as
\beq
\gam\equiv {\eps(1-6\xi)\over 6\xi \mu^2}\,.
\eeq
We note that the $(1-6\xi)$ factor appears in all the graviscalar
couplings.

In addition, it is useful to define
\beq
\rn\equiv {\gam\over \eps}(\mn^2-\mu^2+i\rho)\,.
\eeq
Then, the full $WWh'$ and $WWs'$ couplings take the form
\bea
g_{WWh'}&=&\left(1-\half \sumn \eps^2 {1\over (\mn^2-\mu^2+i\rho)^2}\right)
g_{WWh}+\sumn \left({-\eps\over \mu^2-\mn^2-i\rho}\right) g_{WW\sn}\nn\\
&=&\left(1-\half\sumn {\eps^2\over (\mn^2-\mu^2+i\rho)}[1-2\rn]\right)g_{WWh}
\eea
and, similarly,
\bea
g_{WW\sn'}&=& {\eps\over \mu^2-i\rho-\mn^2}g_{WWh}+\left( 1-\half {\eps^2\over
    (\mn^2-\mu^2+i\rho)^2}\right)\gam g_{WWh}
=-{\eps\over \mn^2+i\rho-\mu^2}[1-\rn]g_{WWh}\,,
\eea
where we dropped the $\eps^2$ term in the big parenthesis since $\gam$
is already of order $\eps$. We can now correct the computation
we did in Eq.~(\ref{stepequation}). Following the same type of
procedure we obtain (dropping the common $g_{WWh}$):
{\footnotesize
\bea
&&\cala_{WW\to WW}\nn\\
&\sim &\frac i {p^2 -\mu^2+i\rho+\sumn {\eps^2\over m^2-\mu^2+i\rho}}
\left(1-\half\sumn\frac {\eps^2}{(\mn^2-\mu^2+i\rho)^2}[1-2\rn]\right)^2 
+\sumn \frac i {p^2 -\mn^2-{\eps^2\over \mn^2-\mu^2+i\rho}}
\left(\frac {-\eps}{\mn^2-\mu^2+i\rho}[1-\rn]\right)^2\nn\\
&=& \frac i {p^2 -\mu^2+i\rho-\sumn {\eps^2 \over p^2-\mn^2}\left(1- (p^2-\mu^2+i\rho)
[2{\gam\over \eps}] +(p^2-\mu^2+i\rho)^2
\left({\gam\over\eps}\right)^2\right)
}
\,.
\eea
}
In the on-shell
approximation of $p^2\sim \mu^2$ the corrections to the terms we kept before are of order
\beq
2i\rho{\gam\over\eps}-\rho^2 \left({\gam\over \eps}\right)^2
\sim 2i {\Gamma_h^{SM}\over\mu}{1-6\xi\over 6\xi } -
\left({\Gamma_h^{SM}\over\mu}\right)^2\left({1-6\xi\over 6\xi } \right)^2
\eeq
where we used $\rho=\mu\Gamma_h^{SM}$ (where $\mu$ is our short-hand
notation for $\mh$).
For a light Higgs, $\Gamma_h^{SM}/\mu$ is a very tiny number and this
correction can be neglected.  (Note that for small
$\xi$, $\eps\propto \xi$ so that there is no actual
$\xi$ singularity and the whole effect is simply very small.) Our
numerical results presented in the main body of the paper neglected both this correction and the $F'$
correction discussed earlier.

Let us finally conclude  with the obvious generalizations of Eqs.~(\ref{hdef}) and (\ref{sdef}): 
\begin{equation}
 h\sim N \left [ h'+\sum_{\vec m>0}\frac {\eps} {m_h^2-i\rho-m_{\vec m}^2} 
 s_{\vec m}'\right ]
\label{hath}
\end{equation}
\beq
 s_n= N_{\vec n}\left [ \sn'- \frac {\eps} {m_h^2-i\rho-m_{\vec n}^2}
  h' -\half {\eps^2\over (m_{\vec n}^2-\mh^2+i\rho)}\sum_{\vec
    m\neq\vec n,\vec n>0,\vec m>0} {1\over m_{\vec m}^2-\mh^2+i\rho}s'_{\vec m}\right]
\label{hats}
\eeq
where $h$ and $s_n$ are the original fields before diagonalizing the
Hamiltonian and 
\begin{equation}
N\sim \left [ 1+\sum_{\vec m>0}\frac {\eps^2} {(m_h^2-i\rho-m_{\vec m}^2)^2} 
\right ]^{-1/2}
\label{Ndef}
\end{equation}
\beq
N_{\vec n}=\left [1+\frac {\eps^2} {(m_h^2-i\rho-m_{\vec n}^2)^2} \right
]^{-1/2}\sim 1 +\calo\left({1\over\mpl^2}\right)\,.
\label{Nndef}
\eeq
We emphasize again that Eqs.~(\ref{hath}) and (\ref{hats}) yield a
diagonal, canonically normalized form for the kinetic energy terms
while the mass terms also take a diagonal form:
\bea
&&-\frac 1 2\left [(\mh^2-i\rho)h^2 + \sumn \mn^2{ s}_n^2 - 2\eps  h \sumn { s}_n
\right ]\nn\\
&=&-\frac 1 2 \Bigg [ \left(\mh^2-i\rho+\eps^2\sumn \frac 1{m_h^2-i\rho-\mn^2}\right) h'^2+
\sumn  \left(\mn^2
-\eps^2 \frac {1}  {m_h^2-i\rho-\mn^2}\right)
\sn^{\prime\,2}
\Bigg ]\nn\\
&&+O(\eps^3)\,,
\eea
Let us study the $h'^2$ mass
squared.  We have
\bea
m_{h'}^2&=&\mh^2-i\rho+\eps^2\sumn \frac
  1{m_h^2-i\rho-\mn^2} \nn\\
&=& \mh^2-i\rho +\eps^2\sumn {\mh^2-\mn^2+i\rho\over
  (\mh^2-\mn^2)^2+\rho^2}\nn\\
&\simeq&\mh^2\left(1+{\eps^2\over \mh^2}\Re\left[\sumn \frac
1{m_h^2-i\rho-\mn^2}\right]\right)-i\left(\rho-\eps^2\sumn \pi\del(\mh^2-\mn^2)\right)\,.
\eea
From this result we see that the Higgs mass renormalization is given by
\beq
\mhprime^2\equiv \mh^2 \left(1+{\eps^2\over \mh^2}\Re\left[\sumn \frac
1{m_h^2-i\rho-\mn^2}\right]\right)
\label{mhprimedef}
\eeq 
in agreement with Eq.~(17) of Ref.~\cite{Giudice:2000av}.

\bibliography{hinv_simple_6}

\end{document}